\documentclass[sigconf]{acmart}
\usepackage{soul}

\AtBeginDocument{%
  \providecommand\BibTeX{{%
    \normalfont B\kern-0.5em{\scshape i\kern-0.25em b}\kern-0.8em\TeX}}}

\copyrightyear{2023} 
\acmYear{2023} 
\setcopyright{acmlicensed}\acmConference[CHI '23]{Proceedings of the 2023 CHI Conference on Human Factors in Computing Systems}{April 23--28, 2023}{Hamburg, Germany}
\acmBooktitle{Proceedings of the 2023 CHI Conference on Human Factors in Computing Systems (CHI '23), April 23--28, 2023, Hamburg, Germany}
\acmPrice{15.00}
\acmDOI{10.1145/3544548.3581548}
\acmISBN{978-1-4503-9421-5/23/04}

\begin{document}

\title[Exploring Perspectives of Recruiters and Job Seekers on Recruitment Bias and Automated Tools]{``Finding the Magic Sauce'': Exploring Perspectives of Recruiters and Job Seekers on Recruitment Bias and Automated Tools}

\author{Mitra Lashkari}
\email{mitra.lashkari@polymtl.ca}
\affiliation{%
  \institution{Polytechnique Montreal}
  \city{Montreal}
  \state{QC}
  \country{Canada}
}
\author{Jinghui Cheng}
\email{jinghui.cheng@polymtl.ca}
\affiliation{%
  \institution{Polytechnique Montreal}
  \city{Montreal}
  \state{QC}
  \country{Canada}
}

\begin{abstract}
Automated recruitment tools are proliferating. While having the promise of improving efficiency, various risks, including bias, challenges the potential of these tools. An in-depth understanding of the perceived risk factors and needs from the perspective of both recruiters and job seekers is needed. We address this through an interview study in the high-tech industry to compare and contrast the concerns of these two roles. We found that the importance of clarifying position requirements and assessing candidates as ``whole individuals'' are commonly discussed by both recruiters and job seekers. In contrast, while recruiters tended to be more aware of cognitive bias and desired more tool support during interviews, job seekers voiced more desire towards a healthy candidate-company relationship. Additionally, both roles considered the uncertainty of the current technology capability and reduced human contact as concerns for using automated tools. Based on these results, we provided design implications for automated recruitment tools and related decision-support technologies.
\end{abstract}

\ccsdesc[500]{Human-centered computing~Empirical studies in collaborative and social computing}

\keywords{Hiring, automated recruitment tools, bias, decision making, decision support system}

\maketitle

\section{Introduction}
Recruitment of future employees is an essential activity in any organization. At the same time, the recruitment process can be tedious for both the organizations that try to acquire the most competent candidates and job seekers who aim to get a satisfactory position. Many companies and organizations use automated tools to alleviate the efforts required in the recruitment process. Those tools can support the filtering and selection of candidates~\cite{Freire2021}, conducting virtual interviews~\cite{SUEN201993}, and providing job recommendations for job seekers~\cite{Charleer2019}, to just name a few.

While promising, many risks and challenges are involved in the use of automated recruitment tools. For example, an early adopter of such tools, Amazon, found in 2015 that their machine learning-based tool for matching candidates to job positions was biased toward men~\cite{dastin_2018}. Journalists from MIT Technology Review also questioned the reliability of automated tools for detecting personality traits from voice recordings of answers to open-ended interview questions~\cite{Wall2021}. Moreover, highly automated interviews are found to have posed challenges in ensuring fairness and allowing job seekers to have social contact with the company~\cite{Langer2019}. Overall, recruitment bias and misjudgement are widely acknowledged in automated tools that support this complex decision-making process.

In order to allow automated recruitment tools to fulfill their promises, an in-depth understanding of the perceived risk factors and needs of the users of these tools is necessary. Recruiters are often considered to be the primary users of recruitment tools since they often drive the decision-making process. However, we argue that the design of recruitment tools should also consider the job seekers' perspectives and needs, because of the following two main reasons. First, job recruitment is a two-way choice activity, in which job seekers actively make decisions based on their experiences with the company or organization, mainly through the recruitment process. Any issues or inconveniences caused by recruitment tools will not only give a bad impression of the company to the candidates but also add to the stress and anxiety that the job seekers already experience during the recruitment process. 
Second and more importantly, adopting a feminist HCI~\cite{Bardzell2011} perspective, considering the experiences and needs of the non-dominant group in the recruitment process (i.e., the job seekers) is not only beneficial, but also crucial, to the design of a truly successful and innovative recruitment support tool. HCI researchers and designers should engage the imbalanced power relationship and advocate for the usually marginalized group (i.e., job seekers). In our opinion, this not only should be done in job seeker-facing technologies (e.g.,~\cite{Charleer2019}), but also and perhaps more importantly, needs to be a focus in the traditionally recruiter-facing environment, such as the recruitment tools. 

Adopting this point of view, in this paper we aim to understand the perspectives of both sides of the core stakeholders of the recruitment decision-making process (i.e., recruiters and job seekers) on the important topic of recruitment bias and automation. 
We particularly ask the following research questions from both the recruiters' and the job seekers' perspectives:
\begin{itemize}
    \item RQ1: What factors can potentially lead recruiters to have a bias and misjudgment?
    \item RQ2: What are the best practices for recruiters to address bias or misjudgment?
    \item RQ3: What are the risk factors in automated job recruitment tools that might lead to bias and misjudgment?
    \item RQ4: What are the desired features of automated job recruitment tools?
\end{itemize}

When answering these research questions, we focused on recruitment in the high-tech industry, and thus job seekers who are likely to have received higher education. This decision is made because (1) companies in this industry are (or if not already, likely to be) early adopters of automated job recruitment tools and (2) our familiarity with this industry will allow us better analyze and interpret the data. Thus in this context, we conducted an interview study with ten recruiters who are experienced in recruitment decision-making and eight job seekers who recently underwent a job hunting process.

In the first three research questions, we deliberately focused on issues related to bias and misjudgment. There are various and sometimes conflicting definitions of these terms in the literature~\cite{acciarini2020cognitive}. For the purpose of our study, we consider the most common and straightforward definitions. According to the Oxford English Dictionary, \textbf{bias} is defined as \textit{inclination or prejudice in favor of or against one thing, person, or group compared with another, usually in a way considered to be unfair}, and \textbf{misjudgement} is defined as \textit{the action of forming a wrong opinion or conclusion about someone or something}. In these definitions, both terms carry the connotation of making a mistake in the estimation of someone or something. There are several nuanced differences: (1) the concept of bias often describes the tendency before the estimation (i.e., prejudice) while the concept of misjudgement emphasizes wrong conclusions after the estimation; (2) to have bias, two or more groups are involved while this is not necessary for misjudgement; and (3) bias is connected with the concept of fairness while misjudgement is with correctness (i.e., right or wrong). Considering these definitions, in the context of recruitment, bias describes the recruiters' conscious or non-conscious prejudice for or against a group of candidates, while misjudgement describes the various inaccurate estimations that the recruiters can make when assessing a candidate. We considered both concepts during data analysis to capture a wide scope of aspects related to the potential mistakes made during the recruitment process and with automated tools.

Through a thematic analysis of the interview data, we found that recruiters and job seekers have certain common perspectives related to all four research questions. Particularly, these common concerns, best practices, and technology needs are related to the clarity of position requirements and expectations, as well as the strategies for assessing candidates as ``whole individuals.'' However, while our recruiter participants tended to reflect on the cognitive bias that can appear during recruitment decision-making and expressed more technological needs to support conducting interviews, the job seekers voiced ideals to have a more equal and constructive relationship with the company during the recruitment process. Together, our results indicated that future investigations of automated recruitment support tools should focus on addressing cognitive bias in the recruitment process, supporting team decision-making, facilitating (rather than performing) decision-making, and empowering job seekers while supporting recruiters. Our work provides valuable knowledge for advancing the research and the practice of automated recruitment tools and related decision support systems.
\section{Related Work}
We build our research on the prior literature related to (1) bias, discrimination, and general challenges in the recruitment process, (2) technology's role in supporting the recruitment process, as well as (3) bias and fairness in recruitment-support tools.


\subsection{Bias, discrimination, and general challenges in recruitment}
Many previous studies have identified that bias and discrimination are major issues in the recruitment process. For example, Beattie and Johnson~\cite{Beattie2012} identified various sources and impacts of unconscious bias in recruitment and promotion processes. They argued for the importance of raising awareness and promoting equity in recruitment. Whysall~\cite{Whysall2018} conducted a literature review to identify implicit bias in various recruitment steps, such as resume screening, interviewing, and performance evaluation. They found that the discrimination involved in the recruitment process is ``subtler, deeper routed'' and requires non-conventional approaches to address. Overall, race and gender biases are the most prominent types of recruitment biases that are extensively studied~\cite{Liebkind_Larja_Brylka_2016, Abramo2016}.

Some studies have also focused on investigating ways to address unconscious bias in recruitment. In 2007, Collins~\cite{collins2007tackling} analyzed the impacts of the ``Rooney rule'', introduced in 2002 by the National football league (NFL) to increase the diversity among head coaches. The Rooney rule mandates that every NFL team interview at least one minority for head coaching and front office positions or else there would be a fine. The paper argues that since unconscious bias is unintentional and hard to recognize, solutions such as exhortation, education and protesting only modify the conscious beliefs and leave the unconscious core untouched. Therefore, there is a need of addressing the problem with such hard affirmative action. Echoing this point of view, Williamson et al.~\cite{Williamson2018} also questioned the effectiveness of unconscious bias training in addressing equity in work environments. They argued that the popularity of unconscious bias training makes it to be perceived as a ``silver bullet'' and negatively impacts the implementation of other interventions and actions that may have more sustainable effects.

Previous studies have also focused on identifying various challenges that impact the recruitment process. For example, Rozario et al.~\cite{rozario2019challenges} conducted a survey study with hiring members as well as successful and unsuccessful applicants to understand the ``critical factors'' involved in the recruitment process. As a part of this study, the authors found that unsuccessful applicants were more likely to perceive the impact of bias in interviews than successful applicants and that applicants who perceived bias are more likely to urge improvements in the interview process. Focusing on proposing recruitment technology, Koivunen et al.~\cite{Koivunen2019} dived into the challenges of recruitment decision-making from a ``social matching'' perspective. Through interviews with recruitment decision-makers, the authors proposed a framework of recruitment decision-making stages and discussed challenges in each stage of the framework. They identified, for example, unclear matching requirements and narrow views of individuals' qualities make it challenging to \textit{Search for Expert}. They then proposed several ways that technology can help address the identified recruitment challenges.

Our study extends this body of related work to dissect the challenge of bias and discrimination in the recruitment process. We particularly contextualize this investigation in the prominence of automated job recruitment tools.


\subsection{Technology supporting the recruitment process}
Previous studies have investigated technology support to facilitate many aspects of the recruitment process. Visa et al.~\cite{visa2015new} demonstrated that improving the recruitment process with new technologies is feasible and resource-saving by combining traditional, manual recruitment with automated systems. So{\l}ek-Borowska and Wilczewska~\cite{solek2018new} analyzed both advantages and disadvantages of new technologies for the recruitment process. They found that while technologies may help save money and time, strengthen the image of the company, and reach a broader audience of candidates, they may impede relationship building, candidate integration, and personal data privacy. Regarding the last aspect, Nickel and Schaumburg~\cite{Nickel2004} investigated privacy issues and user trust in online recruitment; they found that the perceived privacy of an online recruitment service has a positive impact on user trust.

Moreover, using technology in recruitment can put a disadvantage on certain groups of job seekers. Several previous studies have focused on understanding and addressing this problem. For example, Dillahunt et al.~\cite{Dillahunt2018, Dillahunt2020} designed and evaluated two tools, for resume feedback and interview practice, respectively, targeting disadvantaged job seekers such as those who are currently unemployed or having health issues. Through this investigation, they provided guidelines and strategies to better support those job seekers. 
Hendry et al.~\cite{hendry2017u} also targeted homeless young people and provided vision and design insights for a platform where youth can find short-term employment in their neighbourhood. 
With the popularity of intelligent systems in the recruitment context, Charleer et al.~\cite{Charleer2019} studied how the design of an interactive dashboard can affect explaining and clarifying the suggestion of recommender systems for job seekers in a mediation session. Their tool design gives job mediators the authority to filter the information on the dashboard when presenting it to job seekers to prevent demotivation; it also provides information through visualization to increase users' trust.

Influenced by this body of literature, we investigate the perception of technology support in the recruitment process from both the recruiters' and the job seekers' perspectives. Through this unique angle, we aim to provide design insights and guidelines to address bias in such technologies that reflect the value and needs of both groups of key stakeholders.

\subsection{Bias and fairness in recruitment-support tools}
There are several studies in the literature that investigated the problem of bias and fairness in recruitment-support tools, including online recruitment platforms and automated tools. For example, Hann{\'a}k et al.~\cite{hannak2017bias} followed the traces of gender and race bias in online freelance marketplaces. Based on data collected from TaskRabbit and Fiverr online marketplaces, they found that there is a significant correlation between the perceived gender and race of the freelancers and the evaluations they got on these websites. 
To understand more about whether the internet has any positive impacts on lessening the bias in hiring decisions, Leung et al.~\cite{Leung2020} conducted a study with 206 Amazon Mechanical Turk participants in a simulated recruiting task. They assigned the task of making hiring decisions for a freelance job to the participants to explore different types of online hiring bias based on gender, race, and attractiveness of the job candidate. Using the discrete choice model to analyze the data, they found that gender, racial, and attractiveness discrimination were all evident in the hiring decision-making process. They also identified that as the number of candidates the participants can choose from increased, the racial bias increased too. Based on these findings, the authors made recommendations for designers to mitigate the bias of online freelance platforms.

Researchers have also explored the bias and fairness issues in automated hiring tools. Peng et al.~\cite{peng2019you} decoupled and investigated the sources of bias in a human-in-the-loop hiring decision-making context. They found that making the list of job candidates gender balanced would result in the correction of bias in various, but not all, professions. The authors claimed that this professional difference root in human decision-making bias and their ``persistent preferences''. They also identified that characteristics of decision makers, such as their gender, have an impact on the bias intensity and its direction. Raghavan et al.~\cite{Raghavan2020} investigated how the commercially available automated candidate-screening tools claimed to address and mitigate recruitment bias. They found that the practices of these tools varied and may have an intricate relationship with anti-discrimination laws and regulations. Similarly, S\'{a}nchez-Monedero et al.~\cite{Sanchez-Monedero2020} examined three automated hiring tools used in the UK to understand how they address and mitigate bias issues. They found that the tools' ability to expose and mitigate hiring bias is sensitive to the social and legal contexts in which the tools are used. Our study draws from this body of previous work to understand the bias-related perceptions of the key stakeholders of recruitment-support tools.



\section{Methods}
We conducted semi-structured interviews with two groups of participants (recruiters and job seekers) to understand their opinions and best practices regarding biases in the recruitment process, as well as their perceptions of automated recruitment tools. The study is approved by the ethics board at the institution of the authors.

\subsection{Participants}

\begin{table*}[t]
\centering
\caption{Experience of the recruiter participants of our interview study}
\label{tab:recruiter-exps}
\begin{tabular}{lp{5cm}p{4cm}p{3.8cm}}
\toprule
\textbf{ID} & \textbf{Company/organization type} & \textbf{Current job title} & \textbf{Years of exp. in recruitment decision-making} \\
\midrule
R1 & HR technology and services & Talent acquisition coordinator & 1 year \\
R2 & HR technology and services & V-level management & 8 years \\
R3 & Global technology company & Talent acquisition partner & 6 years \\
R4 & Power systems solutions & Human resource advisor & 4 years \\
R5 & HR technology and services & Technical director & 7 years \\
R6 & HR technology and services & Human resource administrator & 9 years \\
R7 & Mobility and navigation applications & Product manager & 4 years \\
R8 & HR technology and services & Human resource specialist & Less than 1 year \\
R9 & Video game & Creative director & 7 years \\
R10 & HR technology and services & Human resource specialist & 10 years\\
\bottomrule
\end{tabular}
\end{table*}

\begin{table*}[t]
\centering
\caption{Experience and current position of the job seeker participants of our interview study}
\label{tab:jobseeker-exps}
\begin{tabular}{lp{5cm}p{4cm}p{3.8cm}}
\toprule
\textbf{ID} & \textbf{Company/organization type} & \textbf{Current job title} & \textbf{Duration of the most recent job hunting} \\
\midrule
S1 & Cloud applications and platform & UI/UX designer & 5 months \\
S2 & IT outsourcing & Machine learning engineer & 5 months \\
S3 & Global technology & Full stack developer & 3 months \\
S4 & Online travel shopping & Machine learning scientist & 4 months \\
S5 & Mobility sharing service & Software engineer & 2 months \\
S6 & E-commerce & Mobile developer & 1 months \\
S7 & IT outsourcing & Data scientist & 3 months \\
S8 & Internet applications & Data programmer & 2 months\\
\bottomrule
\end{tabular}
\end{table*}

Our participants were recruited from our industry connections, advertisements on LinkedIn, and our personal contacts. In total, ten recruiters (five females and five males) and eight job seekers (three females and five males) participated in our interview study. Our recruiter participants included those who have experience taking part in the recruitment decision-making process, such as human resource specialists, talent acquisition specialists, and managers who had experience making hiring decisions. On average, they had 5.7 years of experience involved in the recruitment process, with ages ranging from 30 to 45 years. Table~\ref{tab:recruiter-exps} summarizes the characteristics of our recruiter participants. 
Our job seeker participants included those who had recent experience applying for job positions. All participants had experience in job hunting during the past two years before the interview, with ages ranging from 25 to 35 years. Table~\ref{tab:jobseeker-exps} summarizes the characteristics of those participants.

\subsection{Data Collection}
The interviews were done from May 2021 to May 2022. The first author conducted all the interviews online through Skype, Zoom, or Microsoft Teams based on each participant's preference. Each interview lasted about 60 minutes. The interviews were video recorded and later fully transcribed.

The interviews with recruiters included the following parts:

\begin{itemize}
\item {\itshape Introduction and warm-up}: We asked the participants to talk about their job titles, responsibilities, and experiences.
\item {\itshape Recruiting practices}: We asked the participants about their typical recruiting process, positive and negative experiences, and common challenges. We investigated their approaches to examining CVs and managing interviews, as well as their expectations of an ideal process.
\item {\itshape Bias}: We asked participants about their awareness and opinions of bias in the recruitment process. We also asked questions regarding how they dealt with bias and misjudgement as well as how they think these elements can have an impact on hiring decisions.
\item {\itshape Automated tools and technologies to support recruiting}: We asked them about the tools they currently use, the effectiveness of these tools, and the challenges they faced while using them. We then asked about their opinion, including the perceived risks and benefits, of some features of recruitment supporting tools. Particularly, we mentioned three features as probes to allow participants to have a concrete perception of automated recruitment tools: (1) an automatic candidate filtering feature, (2) an intelligent dashboard that supports the interview process, and (3) an avatar or a bot that automatically conducts the interview. Finally, we asked about how they think automated tools and technologies can help address unintended bias in the recruitment process.
\end{itemize}

Interviews with the job seekers had the following structure:

\begin{itemize}
\item {\itshape Introduction and warm-up}: We focused on asking about their job title and responsibilities, their target positions in their last job hunting, and their general job hunting experience. 
\item {\itshape Job interview experience}: We probed about their experience in job interviews, the challenges they had, the questions they have been often asked, and their expectation of an ideal job interview.
\item {\itshape Bias}: We then talked about their observed and perceived bias and misjudgment in their hiring process and the efforts they put into avoiding it.
\item {\itshape Automated tools and technologies to support recruiting}: We asked them about their experience using automated tools in the job hunting process. We then asked them to discuss the perceived benefits, risks, and challenges of using automated tools in recruitment. We also asked about their opinions, including the perceived risks and benefits, about some features of a recruitment support tool; we used the same three features in the recruiter interviews as probes. Finally, we asked them about their opinions on how automated techniques can help address recruitment bias.
\end{itemize}

\subsection{Data analysis}
We used a thematic analysis approach~\cite{Vaismoradi2013} to derive themes from the transcribed interview data, with the help of the ATLAS.ti software. We performed the analysis for the recruiters' and the job seekers' interviews separately. For each participant group, we first inductively coded the interviews; each code is under a specific structure based on the subject of the code and the research question they address. We then categorized the codes through an affinity diagram activity. Finally, we found common themes in the participants' comments. Throughout this process, all authors continuously discussed and iteratively refined the codes, categories, and themes. 

\section{Findings}
In this section, we present the results of qualitative inductive analysis of our interviews with recruiters and job seekers. The results are organized based on the four research questions, which were answered from the perspective of both groups of participants.

\subsection{RQ1: What factors can potentially lead recruiters to have a bias and misjudgment?}
\subsubsection{From the Recruiters' Perspective}
We identified the following themes in the factors our recruiter participants mentioned that can affect bias and misjudgment during recruitment. 

\textbf{Human nature of recruiters}:  When discussing the factors that lead to bias, participants frequently mentioned that personal likings and preferences play a role in recruiters' opinions towards the job candidates. For example, some recruiters said that they like to see certain characteristics in CVs, such as being direct, funny, or result-oriented. For example, R3 mentioned: ``\emph{ I will say I am a stickler for those who don't like spelling mistakes when I see them in a resume. So if I had a clean, direct, concise, funny resume in front of me, that would be my ideal process.}'' In certain cases, personal likings of decision-makers can have very strong impacts, as R6 said: ``\emph{I might find that this candidate is great, but the hiring manager pushes back on us and says, `I don't really like this person personally,' or something, or they might bring their own biases towards the recruitment process.}'' Episodes of experiences can also affect recruiters' judgement. For example, participants noted that lying or disrespectful candidates can be bad experiences that affect their mindset about certain candidates. The recruiters explained the challenges to change their attitude and approach toward certain candidates after experiencing dishonest behavior in their practices.
    
\textbf{Lack of domain expertise for recruiters}: Recruiters considered their role as an intermediary between the job seekers and the hiring managers and they frequently described how not having experience in the field they are hiring for can leave room for doubt about whether they chose the right person or not. For example, when asked about their hiring process, R4 said, ``\emph{In the beginning, it was more of trial-and-error, I'll bring someone, and I will not be sure if that person will fit with the manager... I have a really broad understanding of the job, but what is it on a day-to-day basis and the challenges, I won't really know.}''
  
\textbf{Contents and presentation of CVs do not fully represent candidates}: Our findings show that participants believed CVs could not represent all the skills and qualities of a candidate. Participants said that although they usually go through a time-consuming, sometimes even obsessive, process of filtering the CVs to select the best candidates for a role, it is normal that a number of these candidates do not fit for the role or the company after meeting them. For example, R2 mentioned: ``\emph{Sometimes you're going to have candidates who really look good on paper and then when you meet with them you see that there's no fit.}'' Some participants mentioned that focusing too much on indicators such as GPA can lead to losing good participants since academic performance does not necessarily bring in all the skills needed for a role. A candidate's CV is how they present themselves to the company. We found that every aspect of it, from the information presented to its style, can affect the recruiter's impression of candidates. Furthermore, time gaps on CVs, having an insufficient amount of details and data, or on the other hand, having too many details, all make the filtering more challenging and bias-prone for the recruiters. As R5 said, ``\emph{I appreciate having a high-level technical overview about the technologies a candidate worked within their job description, but if it gets too detailed, it becomes a keyword bomb. It's something that it's a more negative signal for me than a positive one.}''

\textbf{Time limit}: All participants considered the limit of time as a challenge in their job. Due to the high volume of resumes to evaluate and the need to find the most suitable candidate for the position as soon as possible, they can not spend more than a few seconds to minutes on each resume. As R6 said, ``\emph{I'll go through the list. I'll take about 12 to 20 seconds per person going through their resumes because it has to be fast. I have so many open positions that I can't spend more than two or three minutes on one candidate.}'' Based on what participants declared, the pressure of losing the candidates to competitors is another reason for rushing the process. In some cases, it would actually increase the chance of losing a fit candidate. The limited time spent on each candidate results in an incomplete understanding of the candidates, which in turn leads to bias and misjudgement.

\textbf{Cultural differences}: According to our participants, unfamiliarity with cultural differences brings about room for misjudgments and bias. Our recruiter participants talked about the interaction and communication patterns between people from the same culture or nationality and how failing to learn about these patterns could lead to misjudgments. For instance, when talking about the bias in different stages of the recruiting process, R7 explained, ``\emph{I find that Western people would display curiosity in a certain way. It's easy for me to screen out -- Like if you give me only Canadian resumes or... French resumes, I kind of see, okay, this is a loner, this is a team player, this is passionate, but if you give me an Indian resume, it's much harder to make that difference.}''

\subsubsection{From the Job Seekers' Perspective}
Our job seeker participants discussed various factors that can lead to bias and misjudgement in recruitment decision-making. We organized them into the following themes.

\textbf{Narrow-minded focus on a particular aspect}: The participants believed that focusing too much only on a particular aspect can lead to a bias towards some applicants. Participants mostly mentioned ``experience'' to be the aspect that sometimes recruiters and hiring managers rely on strictly to filter the candidates. They claimed that using this approach might lead to ignoring the big picture and individual potential. In this regard, S2 said ``\emph{A lot of times, someone who may not have that -- for example, a number of years of experience. I don't understand why it should matter... It's all about problem-solving, at least in software development. The only metric for hiring someone should be problem-solving.}'' 

\textbf{Not considering individual differences}: Participants also believed that the company should take individuals' differences into consideration when judging them. They mentioned the personality and ability of people as factors that might affect the outcome of an interview for the applicants. As S2 put, ``\emph{... and then soft skills I feel like through the questions they ask in the interview it's kind of easy to detect ... the people who are really good at talking and expressing themselves and things like that. I think they can win over people who are more of an introvert and who can't really express themselves but have really good hard skills. Those are things I feel that the recruiters should have in mind.}'' Another reflection of inequality among the candidates is the difference between their familiarity with the automated platforms. As S6 mentioned ``\emph{I remember I had struggled with trying to make sense of the input and output [of the tool used for answering the questions], and then also the state of the question... It was very confusing... I definitely stressed out and like halfway on saying, `Okay, this is going nowhere,' kind of feeling.}''

\textbf{No clear expectation of the position}: The participants pointed out that recruiters not knowing the true requirements and needs for hiring a candidate for a job position can be a hassle in their job search process. For example, S2 said ``\emph{... a lot of bigger companies, they generally know what they want. The job descriptions are a lot clearer and they know. Basically, because they've done a lot of interviews, they generally know how to conduct interviews. But, I would say, smaller companies, they just copy-paste, for the most part, the job description from the bigger companies...}'' Further, some participants talked about their experience of going through multiple rounds of interviews and challenges and being rejected at the last round. They said what made the experience unpleasant to them is that in the end, they felt that the recruiters did not know what they were looking for. S1 talked about this ``\emph{One other bad experience actually been that, I been through all stages of the interviews and a final stage they got back to me and said they realized that they are looking for someone who had more years of experience and they already knew that from the beginning when I was interviewing with them. So, it’s just like they had mixed feelings and they didn't know so they dragged me on towards the end.}''

\textbf{Job descriptions are inaccurate or vague}:  The participants talked about how they find some job descriptions not clear, inaccurate, or unrealistic. They mentioned that the inconsistency in some job descriptions would prevent the applicants have a true understanding of the role and its requirements; it would increase the need for further explanations which usually happen during the interview sessions, when more effort was put in. Some of the participants have also experienced dealing with absurd expectations of candidates to fit into the role. For example, S7 said ``\emph{I've seen some job posts that are so ridiculous. They're looking for 14 years of experience when this field is barely 14 years old. I remember one in my first job search that was looking for a junior-level data scientist with basically every single qualification that you could possibly have as a data scientist. They need to have reasonable expectations of skill levels and know that no one person's going to have every single quality, probably like the technical skill that they're looking for.}''


\subsection{RQ2: What are the best practices for recruiters to address bias or misjudgment?}
\subsubsection{From the Recruiters' Perspective}
Our recruiter participants talked about ways to address bias or misjudgment in recruitment decision-making, which we organized into the following themes.

\textbf{Not relying only on the CV}: The recruiter participants believed a true understanding of a candidate takes place when seeking or asking for more information about the elements they put on their resumes. They felt that judging the candidates based on the impression they have of them before interviewing with them can be unfair and incomplete, regardless the shortlisting is done automatically or manually. Some recruiters said even though they would rely on certain criteria and desired keywords derived from job descriptions to filter candidates, they felt that criteria such as education or experience should not be the only thing to rely on. For example, R6 mentioned the need of looking for more information beyond those listed in CVs: ``\emph{I also think it's unfair ... if I communicate to somebody like, `I need five years of experience, and when someone applies, they don't have that many years of experience... So I always seek more information beyond that.}'' Our participants also mentioned their experiences about hiring the right matches successfully, when those candidates were not good fits for the position on paper. For example, R3 explained as a part of how they created a shortlist of candidates for the interview ``\textit{Sometimes I have individuals who have none of the qualifications the hiring managers first listed and they've hired them because of everything else that became a part of that personal story that they just never thought of including on their resume.}'' In other words, there are candidates who are unqualified on paper but address hiring managers' unsaid needs. The problem, though, is how to identify these candidates. R6 described their strategy for addressing this problem: ``\emph{I think in order to mitigate those biases, it's important to go through external websites like LinkedIn -- I will reach out to candidates that might not have that background. That's how I eliminate the bias... I'm assuming many other recruiters do the same.}'' 

\textbf{Consider individual values and differences}: Related to the previous theme, the recruiters believed that they should take people's differences into account and should focus more on the values each individual can bring to the company. As R6 said: ``\emph{When you're ranking somebody based on their answers, sometimes you just can't put a number to their rankings... It doesn't give you the whole story of the candidate... 
Because sometimes, even though I have rated somebody 80\%, but there's someone who's at 77\%, but that 77\% person actually has better verbal communication or better style of communicating. I can't necessarily give him a higher number just because of his interviewing skills.}'' 
Adapting the interviews according to job candidates is one approach the participants take to include individual differences in their process of decision-making. As an example, R4 shared: ``\emph{Each person is completely different. I mean I've had interviews with people that are so dynamic that they take me off guard -- they just come out and they take almost control, be like let me give you my story... And I'll have other individuals that are much more reserved where I have to ask each question -- all they do is answer that, that I have to go through that more structured interview process.}''

\textbf{Specifying hiring criteria and validating them constantly}: An essential factor from our participants' perspectives was following the steps of setting needed qualifications for the position and then validating these metrics constantly throughout decision-making. Having a set of specified metrics for evaluating a candidate gives a solid map to all the people in the recruiting process to avoid bias and misjudgement based on personal opinions. As R6 said, ``\textit{That's the difficulty when you hire as a team because people have different visions, and then I want to make sure to bring them on the same page... If they could just make sure that we understand what is requested, what will be one part of the skills needed, and can the person do the job. There's really to answer those three basic questions will really help in accelerating the recruiting process.}'' Participants mentioned various strategies that fall under this category, such as multiple rounds of review, verifying causes of rejections in groups, and iteratively evaluating the process. For example, R6 again said, ``\emph{What we do when we have these HR meetings or certain things, you obviously discuss, `Why did this position take a certain time to fill?' Or `What were some of the problems that you faced when you were filling these positions?' For the biases that we do see, we obviously communicate to the director of the HR department, and they, in turn, follow up. They create diversity programs, or they create training modules... So, that works.}''

\textbf{Asking for hiring team members' opinions}: Related to the previous point, the recruiters who participated in our study valued the opinions of diverse members who get involved in the hiring decision-making process. For example, they considered smooth communication with hiring managers to understand the ``\textit{magic sauce of the role}'' (R3) as a crucial step. In general, having multiple people from the recruiters' team in the interview and the filtering stage as teams to cover each other's bias gaps effectively mitigates the related risks. As R2 said, ``\textit{You have two types of biases, you have a positive bias and a negative bias. That's why you have usually several people doing the interview process because you want to mitigate the bias factor.}''
Sometimes, seeking the opinions of others can serve as a way to understand their value and needs, and thus streamline collaboration. For example, R3 mentioned: ``\emph{Usually what I like to do is that, I will present to the hiring team three completely different candidates. For example, their nationality is the same, but they have completely different personalities. They have different ways of expressing themselves. They have different backgrounds, completely. And they have a different flare to how they do their job. So it's up to the hiring team to figure out which one is best suited for the team... Then as soon as you understand which one they actually want, Well, now you understood your whole hiring team from then on.}''

\textbf{Easing up the atmosphere of the interview}: Based on our study, the participants considered easing the interview's atmosphere as a factor to help them know the candidates better. They claimed that easing the candidate's spirit would let them present themselves as the best without feeling intimidated by the recruiters; this would help the recruitment team make a more accurate judgement. For example, R4 said ``\emph{If there are topics to discuss with that person and make them more comfortable, [we will do it]. We can talk about hobbies, for example... Then just to ease the spirit, just to ease the interview down a little bit because it's not easy to be the only person ... when you have like five, six other people in front of you and then it's a little bit intimidating.}''

\textbf{Asking open-ended, transparent, and honest questions}: From our participants' perspective, by asking open-ended questions, the recruiter would be able to get more insights about the candidate based on the way they answer the questions. As R5 said ``\emph{I try to keep the questions open-ended. If you are asking questions that I have the right answer and the candidate simply doesn't know the answer I had in mind -- I think it doesn't help really to understand the candidate. My goal is to make the candidate talk, because we'll get more insights into the professionality of the candidate and find more effective ways to harvest information.}'' They also think that honesty and transparency are important factors. For example, R3 said: ``\emph{Things like red flags, like why are you not able or want to have growth in that particular company. Those questions I'll ask and I'll ask them very directly. Because everybody has their style. But I usually at the start of the interviews being like, I work with honesty, transparency and clarity. And, as long as we can have it on the call we'll have success together...}''

\textbf{Talking openly about bias}: Our participants state that talking about the bias openly and training and educating the recruitment team would pave the way to increase employers' and employees' awareness of the bias and prepare their minds to detect it. For example, R7 said ``\textit{[To deal with my bias,] I go to events to learn about how to be inclusive in recruitment. I read about it... You have how to include trans people in your recruitment. Then you have this for people of color, you have this for women in male-dominated industries... I try also internally to talk about it with my colleagues ... so we can learn from each other as well.}''. Likewise, they believe the more people know about cultural differences, the better they learn about the norms and differences when gathering a diverse team. For example, R4 said, ``\emph{What they told us when we go to school is like, `Oh you shouldn't spend more than 10 minutes in the resume.' But I don't like doing that because ... for some cultures, they were told to write the most that you can in your resume because that represents you.}''

\textbf{Proactively cultivating diversity}: Our participants said that to mitigate bias, companies should consider creating diversity among employees and reaching out to the whole community as their human resources responsibility. For example, R7 said: ``\textit{I try to have some diversity. I've always worked in male-dominated industries. If we have women in the team or if we have people of color in the team, I try to have them interviewing also so the person is not just facing white dudes. Diversity also comes into play before because when I post the job adverts... I make sure I reach out. There is queer groups in tech, there are women groups in tech, and there are some events or networking events, and I know that those candidates won't come to me necessarily, organically, and so I reach out.}'' Our participants said that to mitigate bias, companies should emphasize diversity and encourage themselves to utilize quantifiable measures related to diversity while recruiting. They criticized ``meritocracy'' (R7), the merit-based recruitment strategy that excessively emphasizes the employees' performances over diversity.

\subsubsection{From the Job Seekers' Perspective}
From our job seeker participants' perspective, there are several things the recruiters can do to address bias or misjudgment in their decision-making. We organized these recommended practices into the following themes.

\textbf{Recruit as a team}: Our job seeker participants mentioned that the recruitment decision-making should be done by a team, instead of controlled by a few people, to avoid personal bias. For example, S2 discussed this point: ``\textit{Let's say, ten people are involved in making the decision. If all these ten people are kept in the dark and they only provide that feedback to the recruiter and the recruiter makes a decision finally, that's not a good meter. You want to have some sort of shared knowledge.}'' Participants also voiced the importance of having diversity in the recruitment team. Many times, this recruitment team diversity also gives the impression of the diversity of the company. S7 talked about this aspect: ``\textit{I really noticed what were the genders and backgrounds of the people who were interviewing me... It was just something I noticed because if this is the team you're going to be working with, I am curious, 'Are you all men? Are you all White men? Who was asking me questions? Who was in the position of power to be giving me the go-ahead or not?'}''

\textbf{Ease-up the interview atmosphere}: Our participants explained how the interview environment and atmosphere can affect their performance, especially when doing a technical challenge. Having other people observing their work whether they are doing the challenge in-person or virtually make participants uncomfortable and divert their focus from the important matter. For example, S7 said ``\emph{I had a live coding test in my first job search. I failed miserably at it, and didn't get the job, but anytime I've had a take-home assessment, I've been able to complete it much more solidly, much more confidently... [During the failed coding test] the guy was sitting right next to me. It was too much pressure.}'' Interestingly, participants regarded putting comments on their answers or work during the interview as having a destructive impact on their self-consciousness. For example, S5 shared their experience regarding this: ``\emph{He [the interviewer] asked me to code the entire class of poker game style, like a poker card. I think I was a little bit slow on the coding side. I know all the concepts. I think I understand the design pretty well. ... But on the coding side, for the past year, I wasn't doing Java coding for a while. ... So I needed a little time to brush up on that. He had comments like, `You don't know this?' With a surprising sound... Instantly, my heart just -- After that, nothing went well.}''

\textbf{Focus on role-related questions}: Our participants' opinion on the questions asked in the interview sessions is that it is better for recruiters to avoid repetitious and generic questions during the ``team matching'' interviews, but focus on role-related questions about the candidate's experience. 
Relating to this S2 said: ``\emph{[An ideal interview experience for me would be] basically the behavioral parts, the team match, all those parts, be more meaningful. Really try to get to know someone through their experience, and avoid generally those soundbite questions.}''
Also in terms of technical questions, they preferred to be asked efficient and less time-consuming questions related to the role they are interviewed for. For example, S2 shared ``\emph{
It has to be very similar to the problem that you're trying to solve. ... Like it's good to know data structure to do AI, but they don't necessarily match and vice versa. The best thing that you can do is just, if you have a problem of your own and you're trying to solve or debug, and make a similar question out of it and ask that question. If the person can solve it, that's it, you don't need to look for anything else in terms of technicalities.}''

\textbf{Respecting applicant's time and interests}: Based on our interviews, companies' respect for job candidates’ time limitations and efforts in gaining information about the candidates can help the candidates to have a better experience in their job interview process. It can help reduce the stress and anxiety of the candidates during the process and maintain a positive view of the company in candidates' minds. 
For example, S1 said ``\emph{Peoples who actually listen to what you're saying and ask good follow-up questions that is the best thing the recruiters can do. Putting time into really reading your resume if they have any questions about that area they could ask us and then we could be more specific and let them know. Because sometimes during interviews you don't have enough time to talk about the stuff you've done. They ask one question and you have to get everything out and let them know you are capable of doing things.}''

\textbf{Do not rely on the CV as the only source of information}: Our job seeker participants expressed the importance of presenting themselves aside from what is said in their CVs. For example, S1 talked about this while mentioning their effort accommodating to the companies' mental model: ``\emph{...when explaining my work because I've done the research-based master's. I've basically done two years of research on a project that I focused on. So I tried to put that in focus and let them know that I've worked on a real project.}'' 
Applied to the recruiters' side, the job seeker participants also believed recruiters should put real effort into getting to know their candidates, going beyond what is put in the CVs. 
For example, job shadowing programs have been mentioned as a way of getting to know the job and environment better as well as being evaluated by different people in the team. As S5 said: ``\emph{The first job I had, the interview, the manager had a shadowing program with me for the last round of the interview. ... I was in the company and ... watched over his shoulder. He worked, and then he asked me questions like, `What does this line do?' Or he randomly throws questions at me more like a conversation style. And he leads me to different team members, also the team he worked with. Basically, I met everyone that I was going to work with. I also see from inside the company, I think that was really good. Even just say hi to the team. I think to have more, I guess, eyes and then make a judgment, so it's not biased based on one person.}''

\textbf{Provide clear information about the job}: Our participants mentioned that generally, job seekers need more information on the role they apply to and that some job descriptions can be too ``broad''. The job seeker participants expect companies to give them the chance to gain clear information on the daily job of the role, things to expect in the job, and the company’s culture. As S8 said: ``\emph{If I'm very interested in learning, like someone early in their career, it would be nice in the interview for them to ... give me some information about how the company can also help me, rather than just trying to figure out what I can provide to the company. ... I don't always get that information. The ideal interview I think would also give me information that I didn't previously have, or leave time for me to ask some questions because I often have questions and usually, we run out of time when the interviewer is asking me questions.}''

\textbf{Honest and specific feedback}: Participants noted they put a lot of time into preparing for interviews, participating in the interviews themselves, and performing the technical tasks and challenges. These interviews can also sometimes have many stages, especially in big companies. The participants said having feedback on how they did in the interview process after rejection is what companies can do in return for the time they spent in the process. This way, companies are able to convey a sense of importance to all the candidates, as being rejected without any explanation can be confusing, upsetting, and even discouraging for the candidates. For example, S2 said regarding this matter ``\emph{If you're not getting accepted, the second best thing that you can get is feedback, good feedback that you know this part I have to work on for the future and that's a really good thing. [A company] was the only place where I got meaningful feedback. The other places, usually they don't care that much which is really a question of ethics of it. I think it's wrong. If people, in general, are spending some time preparing to interview with you, the least you can do in exchange for their time is to give them some feedback.}''
Participants also mentioned they prefer to have honest and more specific feedback since auto-responses can be generic and not helpful. As S10 stated ``\emph{Overall, in the two job searches that I've had for this career, anytime I didn't make it ... and I didn't know why, I always asked for feedback. And sometimes I was given feedback. But I have to say that most of the time I felt it was generic, and I didn't really feel like it was genuine feedback that I could take moving forward.}''

\textbf{Transparency}: 
Related to the previous point about receiving feedback, participants added that transparency on the criteria of acceptance can lead to clarity of the reason some candidates get rejected.
Additionally, participants voiced the importance of having transparency when communicating the stage of the process the candidates are at and informing them of the next steps. Not hearing back from recruiters and the uncertainty of waiting time for an answer only add up to a negative recruitment experience. For example, S10 mentioned the matter: ``\emph{I've interviewed with companies that have absolutely horrible communication. They say, `We will get back to you by X date,' and then you don't hear from them and you have to reach out to them and it's super disrespectful. When companies actually get back to you, when they say that they will get back to you, or they'll at least give you an update, that's an ideal part of an interview process for sure.}''

\subsection{RQ3: What are the risk factors in automated tools that might lead to bias and misjudgment?}
\subsubsection{From the Recruiters' Perspective}
Our recruiter participants have used various job recruitment tools that have automated components, such as SmartRecruiters\footnote{https://www.smartrecruiters.com}, LinkedIn Recruiter\footnote{https://business.linkedin.com/talent-solutions/recruiter}, Oracle PeopleSoft\footnote{https://www.oracle.com/ca-en/applications/peoplesoft/}, HackerRank\footnote{https://www.hackerrank.com}, ZipRecruiter\footnote{https://www.ziprecruiter.com}, and Airudi\footnote{https://airudi.com/en/}. Collectively, our recruiter participants are concerned about the following factors related to the risk of bias and misjudgement in automated tools.

\textbf{Current technology capability}: When asked about perceived risks of automated tools in the recruitment process, participants voiced various concerns related to the current technology; those include (1) incorrect recommendations and results, (2) the tool's source of data and data accuracy, (3) bias of the machine learning models used in the tool, and (4) missing good candidates and out-of-the-box thinkers. For example, R4 said, ``\emph{I think what would be one of the big challenges is the amount of data that you can get and the accuracy of the data.}'' R7 also talked about the potential of losing good candidates by using automated tools: ``\emph{What is hard is -- for example, what I said about focusing on someone who is result-driven. Can it be a tool that checks if the way the resume is phrased is results-oriented? Although if you say to me `I'm going to screen out everyone that is not framed this way,' I would be scared to lose good candidates.}'' Some participants mentioned that they would manually examine the recommendations provided by automated tools to prevent any loss.

\textbf{Losing human contact}: Some of our participants said that an essential factor of having an interview with a candidate in an early stage is to have a personal connection. Recruiters mentioned that they are the representative of the company in the candidate's eyes. By remaining in the interview process and communicating directly with the candidates, they are addressing the candidate's needs of having a way to understand and evaluate the company. As R2 said, ``\emph{At the end of the day, recruiting or hiring someone, it's the same process as of selling a platform or buying a car. There's an emotion attached to it. There's a `customer service.' ... It's all about building that relationship. The downside would be having an avatar do too much and lose that feeling... At the end of the day, a candidate is accepting a position because he liked the interview process. He liked the manager he met, he sees himself working in this position with this team.}'' 
Even worse, some participants raised the possibility that candidates might find it an insult if they could not interact directly with people representing the company. For example, R9 said, ``\emph{I think some people, especially in more artistic and creative roles, would see it as quite insulting that their potential future in a company could be defined more by a machine than by real humans.}''.

\textbf{Cognitive load required to use the tool}: A risk factor pointed out the risk of being distracted by a real-time tool that helps recruiters while interviewing. They wanted to ensure the tool would not interfere with recruiters' other tools and focus during the interview. For example, R3 said, ``\emph{So if I've got my WebEx interview going, I would want to have enough space on the screen where I could have it as a user-friendly thing where it's right there, it's not going to interfere with my WebEx space, if you will, and just make it as user friendly as possible and I know that that's easy to say, harder to do.}'' Also, some participants mentioned the time and effort required for learning an automated tool as a challenge for the recruiting team.

\subsubsection{From the Job Seekers' Perspective}
Our job seeker participants mentioned that they have encountered several automated tools during their job hunting processes, such as SmartRecruiter and HackerRank, mainly for filtering candidates; they have also described experiences in which automated techniques were used to conduct assessments (e.g., programming assessment with CoderPad\footnote{https://coderpad.io}). Overall, our job seeker participants had several concerns about automated recruitment tools that we categorized into the following themes.

\textbf{Losing human contact}: A major concern of the candidates was related to losing human contact in the recruitment process when using automated tools. According to our participants, having a company representative during the interviews provides the benefits of connecting to the company and establishing mutual empathy as human beings. Automating parts of the recruitment process where a company representative and a candidate have a chance to make a personal connection would take this opportunity away from the candidates. For example, S5 said: ``\emph{I guess [with automated technologies] there will be less bias but less professional experience from HR as well... Any intermediate person would bring some value into the business as a whole, but if you try to mitigate that, then you may be losing that value from a real person as well.}'' 
Similarly, S6 said ``\emph{Something that I think I've experienced personally is that in the recruitment process, a candidate like me or any candidate for that matter wants to have a personal connection. An automated tool doesn't really provide that... 
For example, if I'm emailing someone, I want the other end to be a human being that really empathizes with me.}''
Based on what participants said, utilizing automated job support tools in some parts would also cause lacking of social feedback. 
The candidates believed when in the interview, facing an automated tool they might never find the chance of having the right impression of how they performed in the interview. For example, S3 shared their thoughts about this matter with us: ``\emph{The reason why is that, when you are doing an interview with the person, when you are answering the questions, there is a lot of indicators you can watch to see if you're doing well. For example, the body language of the person, you will see if he's satisfied at the end with yes or not. If you did not answer correctly, you'll see that he's not convinced or trying to elaborate, so having a real person is really important. Having an avatar, I don't know, just doesn't feel the same way.}''

\textbf{Losing opportunities due to automation}: Based on our interviews, a major concern toward automated job recruitment tools is about losing potential candidates. Our participants argued that since the algorithm for candidate filtering in these tools always relies on some predetermined information, there is a chance it misses other quality information. This is a major risk for them. Some participants mentioned a coping approach that they adopted through the process in order to prevent being filtered out by automated tools. For example, S1 said ``\emph{In the beginning I was expecting to get more interviews ... but it took maybe around three weeks to get my first... And what I found out, later on, was that I need to change my resume for each job description that I wanted to apply for so it would have matching words that are both in my resume and inside the job description. When I started doing that, I started to get more reactions from the recruiters both on LinkedIn and on the other places that I applied.}'' Participants also talked about how the keyword extraction might not work properly with different layouts and styles of resumes (e.g., designers' resumes). Thus, tools relying on such information may not accurate or reliable. Moreover, disadvantaged job seekers can be mistreated by automated tools, as S1 said: ``\textit{Not all job seekers are alike. Like there are disabled people out there so if we're going to rely on something just based on talking or based on facial expressions, it's not going to work for everybody... And then there's always different accents and things like that, which really worries me if the technology is strong enough to accept these kinds of things.}''

\subsection{RQ4: What are the desired features of automated job recruitment tools?}
\subsubsection{From the Recruiters' Perspective}
Our recruiter participants mentioned various features in automated tools that they think would be helpful for them. Those features fall into one of the following categories.

\textbf{Explainable candidates ranking/shortlisting}: The participants talked about a potential candidate ranking and shortlisting feature of automated recruitment tools and voiced their need for explainability and transparency of the decision made in the system. Regarding this, R5 said, ``\emph{What is very important is to have explainable recommendations where we don't need only a number or a ranking of candidates. The system should explain why the candidate is recommended over another.}'' Even when the explainability of individual decision points cannot be achieved, participants still desire to have a general understanding of the automated components of the tool. For example, R7 said, ``\emph{At least, I wouldn't want to see what resumes are filtered out. ... I would like to see how it works. Even if I don't see the algorithm. I would like at least a blog post or something describing how it works.}''

\textbf{Customizable filters}: Our participants talked about their need for a CV filtering feature with the ability to adjust and customize the criteria. They want the tool's flexibility to specify their main requirements, target, and agenda based on the role and then make decisions considering those criteria. For example, when asked about the risks and challenges involved in shortlisting features of the automated tools, R2 said, ``\emph{From one company to another, values and culture are different. A good developer for one company is not necessarily a good developer for another... It's important to be able to have that personalization feature where you can calibrate and adjust the weight of each criterion you're looking for. It would be a danger falling to a generic approach.}''

\textbf{Real-time feedback during interview}: The recruiter participants desired tools that have features that can provide feedback about the interview in real-time, helping them evaluate the candidates and streamline the interview process. For example, R2 mentioned: ``\emph{It would be nice to have a little assistant on the side that identifies or suggests questions that maybe I did not think of asking and maybe scoring the answers to the questions.}''. The participant thought that the tool can provide a second opinion to candidates' answers when ``\textit{the answer they gave us is not the answer we're looking for.}'' 
Furthermore, participants think receiving feedback about how they administrated the interviews with the candidates would help recruiters reflect on their performance and make improvements, as R7 said: ``\emph{I would also like to have some feedback maybe on my tone, all the words I'm using myself just to improve the interviews for me.}''

\textbf{Candidate profile summary}: The participants said they like to see a summary of candidates' information that is derived from the CVs and/or the interviews gone through. They said they would like to see all the information gathered in the different stages of hiring in one place to facilitate an overview and synthesized analysis, even for the stages done automatically. As R5 said: ``\textit{I think it's important that we're not taking the human out of the loop there. The most important feature would be, again, to have a summary of the interaction whether it's chatbots that enable the human recruiter to review the process to make a decision.}'' Participants also have various ideas of how tools can support their analysis to ensure diversity. For example, R7 mentioned: ``\textit{I wonder if it was used in a sense of maximizing diversity within a team, I would be interested if I had a tool that I would have a sense of who do I already have in the team... If I could have a tool that is looking for people that would complement my team, I would be interested.}''

\textbf{Anonymity of the candidates}: Our recruiter participants highlighted the benefits of having an anonymous review of the candidates' profiles to avoid bias. For example, R7 said, ``\emph{For the bias, it would be interesting here to hide maybe names, gender, etc... [These factors] can be brought up at the end to make sure that you have diversity. But it can be hidden from everyone at first.}'' R2 also emphasized the importance of focusing on still fits when screening the CVs, by saying: ``\emph{Some people at a certain time started to put their pictures on resumes. Personally, I did not like that because it gives you a look at how or who the person is. I really prefer having a CV because you look at the skills. When I'm analyzing the CV, I just really look at if it fits with the job description.}'' Generally, participants prefer to have candidate anonymity at the beginning of the profile review. But anonymity itself cannot be considered a silver bullet, as R3 mentioned, ``\textit{bias goes much deeper than just having somebody block out a name.}''

\textbf{Human in control}: Participants generally considered that, with automated tools, ``\textit{it is important that we are not taking the human out of the loop}'' (R5). Part of this desire comes from the lack of confidence in the current capability of the technology, as R9 mentioned, ``\textit{My current understanding is that the level of technology needed to deliver a reliable assessment is many years away from us right now. I think it would have to be handled very carefully, and they would need to be done in conjunction with other aspects... Human intervention would be needed.}'' However, for many participants, human control is an innate factor in the recruitment decision-making process. They are uncomfortable relying solely on automated technology. For example, R4 mentioned, ``\textit{I think you have to have a judgment behind that, some thinking and reflection as a recruiter.}'' R3 also said, ``\textit{The recruiter or the manager need to have that final say, to be like okay I want to actually speak to this person.}''

\textbf{(Automated) transcription and note-taking}: Related to the previous point, our participants also talked about their interest in having an automated note-taking or transcript feature. Also, some participants mentioned the value of having a space to write down their notes to keep all the information in the same place. For example, R3 said ``\emph{So you almost have a... not a spreadsheet but like a questionnaire that you can just fill out quickly as you're speaking to the candidate so you don't have to write it down. 
And if I had a dashboard and I'm speaking to someone, all I need to do is just click off the things that I want to validate. That would be great.}''.

\textbf{Process tracking and interview scheduling}: The participants voiced the needs to be aware of the status of the candidates and the state of the process. Regarding this, R2 said, ``\emph{If that person is not available anymore, I would like to see it in the dashboard. ... If the platform can just remove that person from the data and so I will see how many people withdraw from the position, so I see the urgency of, okay, people are getting really employed in a similar role, so I need to hurry up.}'' Participants also expressed the desires to have technological aid for interview scheduling, especially when they handle many candidates at the same time. For example, R4 talked about this: ``\textit{We spend a lot of time just managing the hours -- what time should I call you, is this time okay with you, or is that time okay with you... In the platform, if we can use it to book directly the candidates instead of sending them validation from my side then [it will be very helpful].}''

\subsubsection{From the Job Seekers' Perspective}
When it comes to the features to be included in automated recruitment tools, our job seeker participants considered the following themes.

\textbf{Mutual understanding between the company and the candidates}: The participants noted the need for both job seekers and recruiters to gain more information about the other party in an automated recruitment tool. On one hand, they suggested having a space where companies can represent themselves and provide more information about the interview stages where the candidate is in. For example, S1 said: ``\emph{I guess on the other side as a job seeker I really wanna know about the company as well. Of course, I wanna apply and get a job sooner, but it really matters to the environment of the company and the culture and all these questions that we job seekers have... So if it's a way it's a tool that would give the company a space to represent themselves and advertise themselves and have employees talk about the company, that would be awesome.}'' Participants also voiced that having more information on the day-to-day job of the role that the candidates apply for, and even further simulating it can benefit the candidates. On the other hand, the idea of the tool deriving more information about the candidates from the internet such as LinkedIn was considered useful from job seekers' perspective. As S5 said ``\emph{[A useful feature of the tool] I think maybe the people's personal project -- GitHub or their blogs or stuff -- then also I guess maybe any recommendation they have from LinkedIn or from their previous colleagues... Maybe, if possible, to get a sense of what they're working on right now as well. Sometimes they are confidential within the company, so it may be hard, but I think it is just really anything to get a sense of how well or how enthusiastic they are about technology in general or what you are hiring them for.}''

\textbf{Honest feedback}: When talking about the useful features of automated job recruitment tools, our participants discussed the crucial matter of feedback and evaluation for both sides again. They expressed how confusing and hard it can be for them not to know how they performed in the process and where the company stands when it comes to choosing them. The participants emphasized knowing the true reason for their rejection can help them in the future. Further, they need to act based on the chance they have with each company. For example, S3 shared ``\emph{I think the bad experiences just you wait and wait. After you pass the interview, and you're not sure if it went well or not, you're not sure if it's better to continue applying or just wait for an offer to come -- the uncertainty of it. The time passes and you become a little stressed out because you didn't receive an offer so far. A little bit of fear kicks in and you start to doubt yourself.}'' On the other hand, feedback on the interviews, process, and the company itself helps answer the candidates' questions about the company. For example, S1 put their opinion on this matter as follows: ``\emph{[I like] having maybe a place that the recruiter can ... after the interview, like put some feedback, a few sentences of feedback, so the job seeker would know... Like, in my idea, I feel like that interview went well, but I never know what the other person is thinking to even improve for the next interview. So having such a place where both the job seeker and recruiter can put a few notes or like a review or something on top of each other's pages, and either private or public -- having something like that [would be useful].}''

\textbf{Anonymity of candidates}: The participants believed a way to mitigate bias using automated tools is their ability to keep the candidates' identities and other factors that might lead to bias hidden. Candidates noted keeping hidden pieces of information such as gender, race, physical appearance, etc. can prevent the formation of judgmental thinking and bias in recruiters' minds. It also helps have all the focus stay on the requirements, skills, and needs of the role. For example, regarding this S2 said ``\emph{Obviously, the second you see the person you have a judgment, you make a judgment about the person. 
I usually don't want to have those. The rounds that you don't need to have a visual cue of the person, it's better to do it with a phone or Zoom without video basically to avoid those kinds of visual biases. Generally avoid all the other sort of stream of information that you can get from the person, and only focus on things that you need in that round. Let's say in the round that is talking about salary and all that stuff, it doesn't need to be a Zoom call; it can be on the phone. Or the round that is technical coding, only coding and nothing else. Only focus on the subject that you're interested in.}''

\textbf{Support thoroughness in decision making}: Our job seeker participants thought that having a tool for calibrating various recruitment criteria among all candidates will be beneficial for streamlining the decision-making process and preventing qualified candidates to be accidentally filtered out. Regarding this, S5 said: ``\textit{I think it will be helpful to have a platform to put all candidates together, so the recruiter will have an overview of everyone, rather than just not opening a link, not seeing your resume at all. Then to have maybe the top criteria that they're looking for list at the top, and then to be clear, your proficiency in each.}'' Similarly, some participants mentioned the importance of supporting knowledge sharing among the members of the recruitment team, in order to ensure a comprehensive view when making recruitment decisions. S2 discussed this aspect by saying: ``\textit{There has to be some level of sharing between those people that are making that decision, and what is that level of sharing has to be very clear.}''

\section{Discussion}
In this paper, we explored the essential factors related to recruitment bias and automated recruitment tools from the perspective of two key stakeholders involved in this process, i.e., recruiters and job seekers. This investigation is particularly conducted in the context of the high-tech industry. While gender and race are the main sources of bias and misjudgement in recruitment, our participants also discussed other factors such as the physical appearance of the candidate, the presentation of the CV, and the recruitment culture in this industry that overly focuses on a particular aspect like years of experience. Overall, our findings highlighted the major concerns, best practices, and technology needs of these two groups of users. In the following sections, we first compare and contrast the perspectives of these two groups. Then we discuss the implications of our findings to the design of automated recruitment tools, as well as the limitations of our current study. 

\subsection{Synthesizing the Key Results}
\begin{figure*}[ht]
\centering
\includegraphics[width=\linewidth]{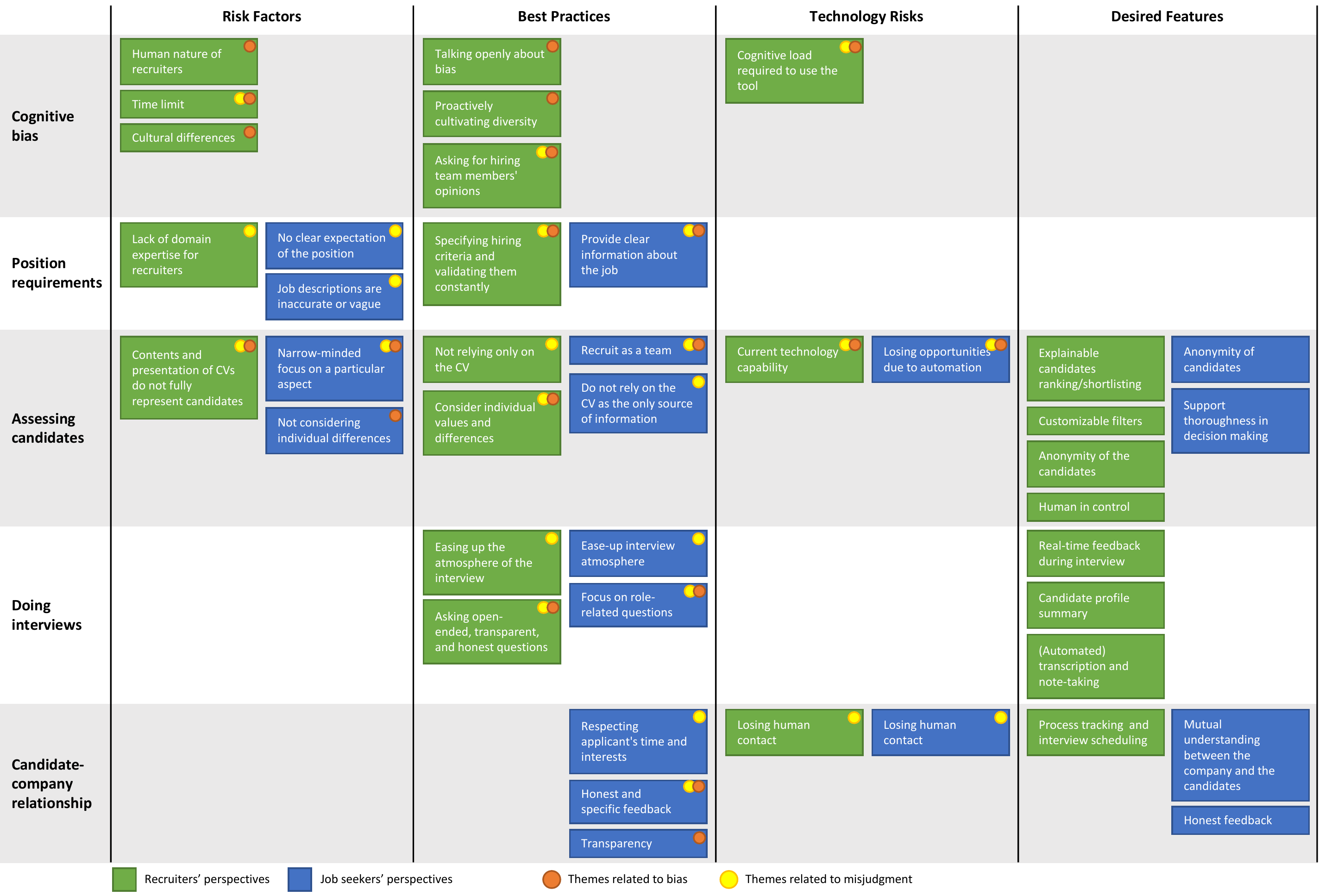}
\caption{Summary of the themes we identified from both the recruiters' and the job seekers' perspectives along five recruitment decision-making aspects. For the first three categories, the orange dots indicate themes related to bias and the yellow dots indicate those related to misjudgement.}
\Description{The figure has five rows, labelled with cognitive bias, position requirements, assessing candidates, doing interviews, and candidate-company relationship, respectively. All themes reported in the results section are put in one of the rows, in the fashion described in the Synthesizing the Key Results section.}
\label{fig:code-summary}
\end{figure*}

On all four research questions, we found that while there are common concerns, suggestions, and desires from the recruiter and job seeker groups, these two groups also provided their unique perspectives on various elements. To better compare and contrast these perspectives, we further organized the themes identified in our findings into five aspects, related to (1) \textit{cognitive bias} in the decision-making process, (2) \textit{position requirements} of the recruiting company, (3) activities conducted for \textit{assessing candidates}, (4) the process of \textit{doing interviews}, and (5) \textit{candidate-company relationship}. We found that along these aspects, our participants discussed issues concerning both bias (i.e., recruiters' systematic prejudice for or against a job seeker group) and misjudgement (i.e., recruiters' inaccurate estimation when assessing a candidate). Figure~\ref{fig:code-summary} summarizes the themes we identified from both the recruiters' and the job seekers' perspectives on these five aspects of recruitment decision-making; for the first three categories, we also indicated the themes related to the concept of bias and misjudgement in the figure. In the rest of this section, we discuss the implications of our findings along the five recruitment decision-making aspects.

\textbf{Cognitive bias}: This phenomenon can be defined as ``cases in which human cognition reliably produces representations that are systematically distorted compared to some aspect of objective reality''~\cite{Haselton2015}. While our job seeker participants did not focus on this aspect, the recruiters seemed to have been well aware of the risks of cognitive bias, discussing many concerns and providing suggestions to help address them. Particularly, the recruiters have touched on all three types of biases in Haselton et al.'s vocabulary~\cite{Haselton2015}. First, our participants considered the biases related to the limited time they can afford to review each candidate, resulting in reliance on intuitions and ``shortcuts'' (i.e., the \textit{heuristic bias}). Second, they discussed biases originating from the human nature of recruiters, who may have particular preferences and inappropriate standards (i.e., the \textit{artifact bias}). And finally, the recruiters mentioned cultural differences as a source of bias, emphasizing that they feel more comfortable assessing candidates from a familiar culture (i.e., the \textit{error management bias}). To address these risks of cognitive biases, our participants made three main suggestions. Interestingly, these suggestions touched on both educative approaches (i.e., talking openly about bias) and actionable interventions (i.e., proactively cultivating diversity and incorporating hiring team members' opinions). These perspectives echo the results of previous studies that educative approaches would only be effective when affirmative actions were put in place~\cite{collins2007tackling, Williamson2018}.

\textbf{Position requirements}: Understanding position requirement concerns the activities of specifying the hiring needs, creating a job description to attract candidates, and establishing criteria for matching and selecting candidates. Hiring managers usually list responsibilities and desired skills that meet the role's demands. They can also get help from recruiters to refine them. Our finding shows that regarding this aspect, the recruiters are mainly concerned about their lack of domain expertise, which would leave room for doubt about choosing suitable candidates. What mainly concerns job seekers is the clarity of the expectation of the position and the ambiguity of the job description.
These two sides of concern, although from different perspectives, are aligned with each other, since having domain knowledge would help address gaps and vagueness in the job descriptions and clarify expectations. As a way of addressing the problems in this stage, job seekers proposed companies provide clear information about the job. The recruiter's perspective on this matter is to specify solid criteria as job qualifications to evaluate the candidates and at the same time work with a hiring team to gain domain knowledge and clarify position requirements. Creating concrete criteria and evaluating these requirements and qualifications over time would lessen the chance of wrong match-making. These concerns and suggestions are also aligned with the previous results that clarifying the hiring goal would help make optimal matches and hiring decisions~\cite{Koivunen2019}.

\textbf{Assessing candidates}: Throughout the recruitment process, it is the recruiters' job to evaluate and compare the candidates based on the position requirements and candidates' qualifications. Previous studies identified that the candidate pools are often heterogeneous and from assessing candidates' points of view, what might prevent the right assessment from happening is being unaware of alternatives and the inability to accurately process all the information~\cite{Koivunen2019, Chua2020}. Our findings show that the recruiters are aware of this and believed that resumes often do not fully represent the applicants. As a way to address this issue, they would consider CV as only a part of the information source. Reaching out to the candidates who might not have the background, looking for more information online, and asking for more information from the applicants are among the practices the recruiters suggested to gain more knowledge about their applicants. It is true that the more information they gain on the candidates, the more parameters will be included in the decision-making. This might make the process more consuming and intricate. Our recruiter participants mentioned another best practice to address this intricacy; that is to focus on the individuals’ values, considering their differences, and knowing their stories. For the aspect of candidate assessment, the job seeker's concerns aligned with those of the recruiters. Particularly, the junior applicants believed that their background is not the only factor that matters, and they feel the need to present themselves more. Reflecting back to the previous aspect related to position requirements, specifying hiring criteria and their importance can be useful to prevent the possible further complexity of the decision-making process for the recruiters. For job seekers, on the other hand, having more information about the job requirements can help them provide more relevant information about themselves during the application process. Moreover, regarding the risks of using technology for assessing candidates, the recruiters and job seekers both voiced concerns about the capability of the current technology in incorporating all the complexity and providing a correct and fair assessment of the candidates.



\textbf{Doing interviews}: Interview is an important aspect in any recruitment decision-making, for both the companies and the candidates to have a better assessment of the other party. Typically, based on the information given by our participants, recruiters conduct a preliminary interview with each applicant (which can take place over the phone). Following that, there are several rounds of interviews with a greater technical focus. These interviews are conducted by the hiring manager or members of the team that is recruiting. In our study, both recruiters and job seekers have expressed the need for an interview environment that is less stressful for candidates. The recruiters stated that they prefer their interviews to resemble a conversation in which both parties can feel at ease and express their points of view. Concerning this issue, job-seekers reported that the atmosphere and environment of the interview affect their performance, especially in technical interviews. With an interview atmosphere that puts each candidate at ease, allowing them to individually achieve their best performance, a fair and equitable assessment can be achieved. Regarding the questions to be asked during the interviews, the recruiters emphasized the importance of including open-ended questions in the interview to allow acquiring more spontaneous and honest information from the candidates. On the other hand, however, the candidates wished for a greater number of role-specific questions as opposed to general and standard questions. This difference in the preferred interview style poses an interesting challenge for automated technologies (e.g., chatbots or avatars) to support the interview process. Additionally, both recruiters and job seekers have expressed their concern about losing human contact when such kind of technology is used. They considered that the connection and the interaction with the other party are important ways to establish mutual understanding and trust, leading to a better assessment and decision-making.

\textbf{Candidate-company relationship}: The recruitment process is often the first step in building the relationship between a company and a future employee. Fostering a constructive candidate-company relationship is also important for building the ``word of mouth'' of the company among potential candidates. Interestingly, based on our results, the job seekers seemed to have more opinions concerning this aspect than the recruiters. While keeping human contact is still an important factor mentioned by both job seeker and recruiter participants, our job seeker participants voiced various very specific suggestions and needs that they wish the companies could satisfy. In the recruitment process, the recruiters are often perceived to be in a position that has more power~\cite{Haugaard2021}. Our job candidates suggested ways that challenge this perception. Particularly, they wished that mutual respect between candidates and the company can be established, that honest and specific feedback can be provided to the candidates for them to understand the decision-making rationale, and that the decision-making process is made transparent to the candidates. Previous work has shown that experiencing power is associated with over-confidence, thus bias and misjudgement, in decision-making~\cite{Nathanael2012}. Focusing on a more constructive and healthy candidate-company relationship by challenging and balancing the existing power dynamic between the two parties can be a viable way to address recruitment bias and misjudgement.


\subsection{Design Implications}
Based on our results and the main aspects of recruitment decision-making that our participants discussed during the interview, the design of recruitment technologies and automated recruitment tools should consider the following elements.

\textbf{Addressing cognitive bias.} Our results revealed that the recruiters are aware of and actively addressing different types of cognitive bias in their work. However, they did not seem to have ideas about how tools can support this consideration. The HCI community has long been investigating the phenomenon of cognitive bias and technology and techniques for addressing it (e.g.,~\cite{Wall2019, Zhang2015}). This knowledge can be leveraged to support recruiters ameliorate their process. However, the current design literature seems to focus on actionable interventions but overlooks educative factors. Our results suggest that both approaches need to be considered for more holistic support.

\textbf{Supporting team decision-making.} Recruitment decision-making relies on teamwork. Our results indicated that involving the expertise and opinions of the recruitment team members can help clarify position requirements and alleviate personal biases; both recruiters and job seekers valued this aspect as a best practice. As a result, recruitment decision-making tools should support this type of collaboration, allowing multiple roles to contribute to the process. Moreover, such tools need to address the intricate issue of ``in-group bias'' that may appear in group decision-making~\cite{CHATMAN2001267}. This is an element that our recruiters discussed when they talk about cognitive biases in general and diversity in particular.

\textbf{Facilitating rather than performing decision-making.} When discussing desired features of automated recruitment-support tools, our participants discussed many themes related to human control in the candidate assessment and interview activities. Not surprisingly, explainability, customizability, and general human control are frequently discussed by our participants when considering tools that automate these activities. Interestingly, our participants have also emphasized ideas for automated technologies used to facilitate, rather than perform, decision-making, such as providing real-time feedback to recruiters, summarizing candidate profiles, and synthesizing candidate information from various sources. Moreover, our participants are generally worried that automation would take away human contact, an important factor that fosters the candidate-company relationship and strengthens decision-making. Thus the concept of automated recruitment tools needs to be broadened and emphasize more facilitating technologies, rather than technologies that directly perform decision-making.

\textbf{Empowering job seekers.} Our job seeker participants often voiced concerns that they feel outpowered in the recruitment process. They desired to have a more equal position with the company, discussing as best practices to obtain more detailed information about the job they apply for and to receive honest and constructive feedback about their performance and qualification. Thus automated recruitment tools should incorporate this perspective and provide features that not only support recruiters but also empower job seekers during the recruitment process. For example, technologies can be developed for job seekers to query information about a specific position. Also, personalized feedback can be generated with the help of automated techniques to allow job seekers better understand the decision-making process.




\subsection{Limitations and Future Work}
There are several limitations of our current study. First, we were only able to interview ten recruiters and eight job seekers. Thus, the generalizability of our findings needs to be examined by future studies with larger samples. However, since our recruitment was terminated by data saturation, we believe that our results represent the major concerns of our target users. Second, most of our recruiter and job seeker participants are from North America. Also, all our job seeker participants have received higher education and most of our recruiter participants are from high-tech companies. Thus, our findings are contextualized in the highly educated population in the North American culture and the job market. This limitation is introduced by the physical location and social connection of the authors. Additionally, this culture and job market are most familiar to us so that we can better analyze the data and interpret the results. However, the issues investigated in this study need to be expanded by future work with a global, more inclusive, view.

Additionally, we adopted the interview research method to explore participants' experiences and perceptions regarding recruitment bias and automated tools. This approach, however, focuses on reflection, which may not be the real practice; it also does not reflect the frequency or nature of bias or misjudgement in actual recruitment processes. While we believe this reflection by the participants provided important information to answer our research questions, future studies using observational methods and analysis of recruitment data, including data generated through automated systems, are needed to further explore the related topics, especially those concerning unintended or unconscious bias. Moreover, in this study we focused on an empirical understanding of the perceptions of the main stakeholders of the recruitment process, but did not study the actual technologies and tools used for recruitment. We argue that understanding the users is a necessary step before any successful technology can be developed, especially when this empirical knowledge does not exist, and that our results provide important insights to inform such technologies. However, future work is obviously needed to investigate the design and implementation of these technologies and tools.

\section{Conclusion}
Automated recruitment tools are proliferating but suffer many challenges and risks that prevent them to fulfill their promises. An in-depth understanding of the perspectives of the main stakeholders, i.e., both recruiters and job seekers, around recruitment bias and automation is needed for making breakthroughs in creating tools that can genuinely support and improve the recruitment process. We targeted this issue and conducted an interview study with ten recruiters and eight job seekers to understand their perceived risk factors related to bias and misjudgement in recruitment decision-making, their suggestions of best practices for addressing these risk factors, their concerns related to automated tools, as well as their desired features in those tools. We found that our participants collectively touched on various aspects related to recruitment bias and automation. The perspectives of recruiters and job seekers are aligned on certain aspects, such as those related to the importance of clear position requirements, the uncertainties on the current technology capacity, and the concerns about losing human contact when automated tools are used. However, we found that the recruiters seemed to be more aware of cognitive bias during recruitment decision-making, while the job seekers voiced more concerns related to establishing a constructive candidate-company relationship. Our results indicated that to be practically usable, automated recruitment tools should focus on addressing cognitive bias in the recruitment process, supporting team decision-making, facilitating (rather than performing) decision-making, and empowering job seekers while supporting recruiters. Our work provides valuable information for researchers and practitioners who focus on automated recruitment tools and other related decision support systems.

\begin{acks}
We thank our participants for their time and thoughtful answers, as well as our anonymous reviewers for helping us improve the paper. We also thank Amanda Arciero from Airudi Inc. for her support and early feedback to our work. This research is partially supported by the Canada Research Chairs program (\grantnum{CRC2}{CRC-2021-00076}) and \grantsponsor{Mitacs}{Mitacs}{https://www.mitacs.ca} (\grantnum{Mitacs}{IT15736}).
\end{acks}

\bibliographystyle{ACM-Reference-Format}
\bibliography{references}
\end{document}